\documentclass[aps,showpacs,twocolumn]{revtex4}
\usepackage{amsfonts}
\usepackage{amsmath}
\usepackage{amssymb}
\usepackage{graphicx}
\usepackage{bm}

\input{tcilatex}

\begin{document}

\title{Transport in charged colloids driven by thermoelectricity}
\author{Alois W\"{u}rger}
\affiliation{CPMOH, Universit\'{e} Bordeaux 1 \& CNRS, 351 cours de la Lib\'{e}ration,
33405 Talence, France}

\begin{abstract}
We study the thermal diffusion coefficient $D_{T}$ of a charged colloid in a
temperature gradient, and find that it is to a large extent determined by
the thermoelectric response of the electrolyte solution. The thermally
induced salinity gradient leads in general to a strong increase with
temperature. The difference of the heat of transport of co-ions and
counterions gives rise to a thermoelectric field that drives the colloid to
the cold or to the warm, depending on the sign of its charge. Our results
provide an explanation for recent experimental findings on thermophoresis in
colloidal suspensions.
\end{abstract}

\maketitle

\textit{Introduction. }Colloidal suspensions in a non-uniform electrolyte
show a rich and surprising transport behavior. Upon applying an electric
field\ or a chemical or thermal gradient on a macromolecular dispersion, one
observes migration of its components and a non-uniform distribution in the
stationary state. The physical mechanisms of electrophoresis and
diffusiophoresis are well understood \cite{Duk74,And89} and widely used in
biotechnology and microfluidic applications \cite{Vio00,Sto04}.

The situation is less clear concerning transport driven by a thermal
gradient. There is no complete description for the underlying physical
forces, and even the sign of the thermophoretic mobility lacks a rationale
so far. It had been known for a while that in some colloidal suspensions the
particles move to the cold, and in others to the warm, corresponding to a
positive and negative Soret effect, respectively \cite{Wie01,Dem04,Pia02}.
Recent experiments on aqueous solutions of lysozyme protein \cite%
{Iac03,Put07}, polystyrene beads \cite{Iac06,Put05,Put07,Bra08}, micelles 
\cite{Iac06}, DNA \cite{Duh06a}, and Ludox particles \cite{Nin08} revealed a
surprisingly similar temperature dependence in the range $T=0...80$ $%
{{}^\circ}%
$C. In all cases, an inverse Soret effect occurs at low $T$, changes sign at
some intermediate value $T^{\ast }$, and seems to saturate above $50$ $%
{{}^\circ}%
$C. On the other hand, a large negative thermophoretic mobility has been
reported for charged latex spheres in a buffered solution at weak acidity
and low salinity \cite{Put05}; adding LiCl or NaCl results in a change of
sign and a transport velocity that depends significantly on the cation.
These features strongly suggest a single mechanism related to the electric
properties of the colloid; the relevance of the thermoelectric effect for
colloidal suspensions has been pointed out recently \cite{Put05}.

A thermal gradient modifies the solute-solvent interactions and drives the
particle at a velocity$\ $\cite{Gro62} 
\begin{equation}
\mathbf{u}=-D_{T}\mathbf{\nabla }T,  \label{2}
\end{equation}%
the coefficient $D_{T}$ being of the order of $\mu $m$^{2}$/Ks. Like any
linear transport coefficient in a viscous fluid, the thermophoretic mobility 
$D_{T}$ has to be evaluated by equilibrating the forces exerted by the
particle on the surrounding fluid with the dissipative stress; the
hydrodynamic treatement is well known, in terms of Stokes' equation with
boundary layer approximation \cite{Ruc81,Mor99,Par04,Fay08,Wue07,Ras07}.

The present paper deals with a non-uniform electrolyte solution. We start by
showing how the Soret effect of the mobile ions leads to a salinity gradient
and a macroscopic thermoelectric field \cite{Eas28,Gut49}. Then we add
charged colloidal particles and study how their thermal diffusion
coefficient $D_{T}$ depends on the electrolyte Soret and thermoelectric
effects.

Consider an electrolyte with monovalent ions of charge $q_{i}=z_{i}e$ and
densities $n_{i}$. The current of each species,%
\begin{equation}
\mathbf{J}_{i}=-D_{i}\left( \mathbf{\nabla }n_{i}+n_{i}\frac{Q_{i}^{\ast }}{%
k_{B}T^{2}}\mathbf{\nabla }T-n_{i}\frac{q_{i}\mathbf{E}_{\infty }}{k_{B}T}%
\right) ,  \label{3}
\end{equation}%
comprises normal diffusion with the Einstein coefficient $D_{i}$, thermal
diffusion with the ionic heat of transport $Q_{i}^{\ast }$, and an
electric-field term. In the stationary state $\mathbf{J}_{i}=0$, one
observes a gradient of the overall electrolyte strength $n_{0}=\frac{1}{2}%
\sum_{i}n_{i}$ and a thermoelectric field $\mathbf{E}_{\infty }$. Both are
well-defined macroscopic quantities, whereas the corresponding charge
separation $\rho _{\infty }=\sum_{i}q_{i}n_{i}$ varies with the inverse
system size and thus is negligible \cite{Eas28,Gut49}. With $\sum_{i}\mathbf{%
J}_{i}=0$ and $\rho _{\infty }\rightarrow 0$ one readily obtains the
salinity gradient 
\begin{equation}
\frac{\mathbf{\nabla }n_{0}}{n_{0}}=-\alpha \frac{\mathbf{\nabla }T}{T},
\label{4}
\end{equation}%
where the reduced Soret coefficient $\alpha $ of the electrolyte solution is
given by the mean heat of transport 
\begin{equation}
\alpha =\sum_{i}\alpha _{i}\frac{n_{i}}{n_{0}},\ \ \ \ \ \ \ \alpha _{i}=%
\frac{Q_{i}^{\ast }}{2k_{B}T}.  \label{10}
\end{equation}%
The thermoelectric field is calculated from the condition of zero electrical
current $\sum_{i}q_{i}\mathbf{J}_{i}=0$; taking $\rho _{\infty }\rightarrow 0
$ one finds \cite{Gut49} 
\begin{equation}
e\mathbf{E}_{\infty }=\delta \alpha k_{B}\mathbf{\nabla }T,  \label{5}
\end{equation}%
with the dimensionless coefficient 
\begin{equation}
\delta \alpha =\sum_{i}z_{i}\alpha _{i}\frac{n_{i}}{n_{0}}.  \label{11}
\end{equation}%
These relations are readily generalzed to higher valencies; for a binary
electrolyte they reduce to $\alpha =\alpha _{+}+\alpha _{-}$\ and $\delta
\alpha =\alpha _{+}-\alpha _{-}$. The origin of the field $\mathbf{E}%
_{\infty }$ is similar to thermoelectricity in metals, where the Seebeck
coefficient is defined as the ratio of induced voltage $\Delta \psi _{\infty
}$ and temperature difference; with typical values $Q_{i}^{\ast }\sim $%
kJ/Mol \cite{Aga89} one finds $\Delta \psi _{\infty }/\Delta T\sim 100$ \ $%
\mu $V/K. In dilute electrolyte solutions, the ionic heat of transport $%
Q^{\ast }$ arises from specific hydration effects \cite{Eas28}; at salt
concentrations beyond a few mMol/l electrostatic interactions become
important and result in intricate dependencies on temperature and salinity 
\cite{Cal73,Cal81,Gae82}.

\textit{Force density. }Now we consider a suspended colloidal particle of
radius $a$ and surface charge density $e\sigma $. Because of the applied
thermal gradient, the permittivity $\varepsilon $ and the Debye length $%
\lambda $ vary along the particle surface, and so do the electric potential $%
\psi $, the field $\mathbf{E}=-\mathbf{\nabla }\psi $, and the ion densities
in the boundary layer. The electric forces lead to a relative velocity $%
\mathbf{v}$ of the charged fluid in the vicinity of the particle, with
additional ion currents $\delta \mathbf{J}_{i}=\delta n_{i}\mathbf{v}-D_{i}(%
\mathbf{\nabla }\delta n_{i}-\delta n_{i}q_{i}\mathbf{E/}k_{B}T)$. Typical
velocities $v\sim \mu $m/s correspond to very small Peclet numbers $\mathrm{%
Pe}=va/D_{i}\ll 1$; thus the convection term $\delta n_{i}\mathbf{v}$ may be
neglected, and the excess ion densities $\delta n_{i}$ in the boundary layer
are given by Poisson-Boltzmann theory, $\delta n_{i}=n_{i}(e^{-q_{i}\psi 
\mathbf{/}k_{B}T}-1)$, where $n_{i}$ describe the pure electrolyte discussed
above. The local charge and excess ion densities read $\rho =-2en_{0}\sinh 
\hat{\psi}$ and $n=2n_{0}(\cosh \hat{\psi}-1)$, with $\hat{\psi}=e\psi
/k_{B}T$.

Thus calculating the thermophoretic mobility reduces to the hydrodynamics in
the charged double layer \cite{And89}.\ The fluid motion is described by
Stokes' equation $\eta \mathbf{\nabla }^{2}\mathbf{v}=\mathbf{\nabla (}%
P_{0}+nk_{B}T)-\mathbf{f}_{0}$, where $\eta $ is the solvent viscosity and $%
P_{0}$ its pressure. The force density $\mathbf{f}_{0}=\rho (\mathbf{E+%
\tilde{E}}_{\infty })-\frac{1}{2}E^{2}\mathbf{\nabla }\varepsilon $ consists
of a charge term with local and macroscopic electric fields, and a
dielectric term \cite{Lan83}. (Typical values $E\sim 10^{7}$\ V/m and $%
E_{\infty }\sim 10^{2}$ V/m imply $E_{\infty }\ll E$.) Rewriting Stokes'
equation as $\eta \mathbf{\nabla }^{2}\mathbf{v}=\mathbf{\nabla }P_{0}-%
\mathbf{f}$ and spelling out the gradients in $\mathbf{f}=\mathbf{f}_{0}-%
\mathbf{\nabla }(nk_{B}T)$, one finds 
\begin{eqnarray}
\mathbf{f} &=&-\left( \rho \psi +nk_{B}T\right) \frac{\mathbf{\nabla }T}{T} 
\notag \\
&&-\frac{E^{2}}{2}\mathbf{\nabla }\varepsilon +nk_{B}T\frac{\mathbf{\nabla }%
n_{0}}{n_{0}}+\rho \mathbf{\tilde{E}}_{\infty }.  \label{1}
\end{eqnarray}%
Note that the force density arises from the slowly varying macroscopic
solvent parameters $T$, \ $\varepsilon $,\ $n_{0}$, and the thermoelectric
field $\mathbf{\tilde{E}}_{\infty }$. The permittivity of water being much
larger then that of the particle, $\varepsilon \gg \varepsilon _{P}$, it
modifies $\mathbf{\tilde{E}}_{\infty }$ close to the interface and, in
particular, enhances the parallel component $\tilde{E}_{\infty }=\frac{3}{2}%
E_{\infty }$.

Following standard arguments \cite{And89}, we solve Stokes' equation in
boundary-layer approximation, that is, for particles larger than the Debye
length $\lambda \ll a$. With local coordinates $x$ and $z$ parallel and
perpendicular to the surface, the force balance in normal direction reads $%
\partial _{z}P_{0}-f_{z}=0$. The normal force vanishes, $f_{z}=0$, implying
constant $P_{0}$. Integrating the equation for the parallel component $\eta
\partial _{z}^{2}v_{x}+f_{x}=0$ with Stokes boundary conditions, one finds
the fluid velocity well beyond the charged layer, 
\begin{equation}
v_{B}=\frac{1}{\eta }\int_{0}^{\infty }dzzf_{x}.  \label{7}
\end{equation}%
In the laboratory frame, the fluid is immobile at infinity, and the particle
moves in the opposite direction with the average boundary velocity, $\mathbf{%
u}=-\left\langle \mathbf{e}_{x}v_{B}\right\rangle $ \cite{And89}.

All forces in (\ref{1}) are proportional to the parallel component $%
T_{x}=\partial _{x}T$ of the thermal gradient. Inserting the reduced Soret
and Seebeck coefficients $\alpha $ and $\delta \alpha $ and rewriting $%
\partial _{x}\varepsilon $ in terms of the logarithmic derivative $\tau
=-d\ln \varepsilon /d\ln T$, we obtain 
\begin{equation}
f_{x}=\left( \frac{\tau \varepsilon E^{2}}{2k_{B}T}-\frac{\rho \psi }{k_{B}T}%
+(\alpha -1)n+\frac{3}{2}\delta \alpha \frac{\rho }{e}\right) k_{B}T_{x}.
\label{6}
\end{equation}%
The contribution in $\alpha $ accounts for the variation of the salinity $%
n_{0}$ along the thermal gradient. The term proportional to $\delta \alpha $
describes the effect of the electric field $E_{\infty }$; it depends on the
sign of the screening cloud and thus of the particle's charge $\sigma $.
Since all contributions in (\ref{6}) are of similar magnitude, the force $%
f_{x}$ and thus the transport coefficient $D_{T}$ may take both signs,
depending on the particle valency and the electrolyte properties. With the
heat of transport $Q^{\ast }$ measured for ions in electrolyte solutions,
both $\alpha $ and $\delta \alpha $\ take values of the order unity that may
be positive or negative. For the case $\alpha =0=\delta \alpha $, as assumed
implicitly in \cite{Ruc81,Par04,Mor99,Fay08,Wue07}, $f_{x}$ is strictly
positive and leads to thermophoretic motion opposite to the thermal
gradient, $D_{T}>0$.

\textit{Thermally driven transport. }We have not yet specified the electric
potential. Gouy-Chapman theory for (almost) flat surfaces gives $\hat{\psi}%
=4 $artanh$(\nu e^{-z/\lambda })$ \cite{Hie97}, where the properties of the
charged particle-fluid interface are condensed in the number $\nu =(1+1/\hat{%
\sigma}^{2})^{\frac{1}{2}}-1/\hat{\sigma}$; the dimensionless coupling
parameter 
\begin{equation}
\hat{\sigma}=2\pi \sigma \lambda \ell _{B}  \label{12}
\end{equation}%
depends on the charge density $\sigma $, the Debye length $\lambda =(8\pi
n_{0}\ell _{B})^{-\frac{1}{2}}$, and the Bjerrum length $\ell
_{B}=e^{2}/(4\pi \varepsilon k_{B}T)$. In the weak-charge limit one readily
recovers the potential in Debye-H\"{u}ckel approximation $\psi =(\lambda
\sigma e/\varepsilon )e^{-z/\lambda }$.

With the explicit expressions for the potential $\psi $, the electric field $%
E=-\partial _{z}\psi $, and the charge and ion densities $\rho $ and $n$,
the integral in (\ref{7}) can be performed analytically, 
\begin{equation}
v_{B}=\frac{1}{\eta }\frac{k_{B}T_{x}}{8\pi \ell _{B}}\hat{C},  \label{22}
\end{equation}%
where the dimensionless quantity 
\begin{equation}
\hat{C}=\hat{\zeta}^{2}+8(\alpha +\tau -3)\ln \cosh \frac{\hat{\zeta}}{4}%
-3\delta \alpha \hat{\zeta}  \label{23}
\end{equation}%
is given as a function of the reduced surface potential $\hat{\zeta}=\hat{%
\psi}(0)$. The relation to the coupling parameter $\hat{\sigma}$ is
established by 
\begin{equation*}
\hat{\zeta}=2\ \text{arsinh\ }\hat{\sigma}.
\end{equation*}%
The most relevant experimental control parameters are the Debye length and
the surface charge density.

\begin{figure}
\includegraphics[width=\columnwidth]{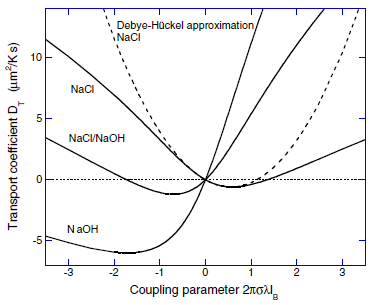}
\caption{Transport coefficient $D_{T}$ for different electrolytes as a function of
the reduced coupling parameter $\hat{\protect\sigma}=2\protect\pi \protect%
\sigma \protect\lambda \ell _{B}$. The full and dashed lines give Eqs. (%
\protect\ref{30}) and (\protect\ref{32}), respectively. With the numbers of
Ref. \protect\cite{Aga89} and Eqs. (\protect\ref{10}) and (\protect\ref{11})
one has $\protect\alpha =0.8$, $\protect\delta \protect\alpha =0.6$ for
NaCl; $\protect\alpha =2.45$, $\protect\delta \protect\alpha =-1.05$ for
equimolar NaCl/NaOH solution; $\protect\alpha =4.1$, $\protect\delta \protect%
\alpha =-2.7$ for NaOH.}
\end{figure}

\begin{table}[b]
\caption{Heat of transport $Q_{i}^{\ast }$ and reduced Soret coefficient $%
\protect\alpha _{i}$ at room temperature for dilute systems. The values $%
Q_{i}^{\ast }$ are taken from Ref. \protect\cite{Aga89}. The parameters $%
\protect\alpha _{i}$ follow from Eq. (\protect\ref{10}). }%
\begin{tabular}{|l|c|c|c|c|c|c|}
\hline
Ion & H$^{+}$ & Li$^{+}$ & K$^{+}$ & Na$^{+}$ & OH$^{-}$ & Cl$^{-}$ \\ \hline
$Q_{i}^{\ast }$ (kJ/Mol) & $13.3$ & $0.53$ & $2.59$ & $3.46$ & $17.2$ & $%
0.53 $ \\ \hline
$\alpha _{i}$ & $2.7$ & $0.1$ & $0.5$ & $0.7$ & $3.4$ & $0.1$ \\ \hline
\end{tabular}%
\end{table}
The transport coefficient $D_{T}$ is obtained by averaging $v_{B}$ over the
orientation of the surface with respect to the applied thermal gradient, $%
\mathbf{u}=-\left\langle v_{B}\mathbf{e}_{x}\right\rangle $ \cite{And89}.
With $\left\langle T_{x}\mathbf{e}_{x}\right\rangle =\frac{2}{3}\mathbf{%
\nabla }T$ and including the factor $\xi =3\kappa _{S}/(2\kappa _{S}+\kappa
_{P})$ accounting for the thermal conductivity ratio of solvent and particle 
\cite{Mor99}, one finds \ 
\begin{equation}
D_{T}=\xi \frac{k_{B}}{12\pi \eta \ell _{B}}\hat{C}.  \label{30}
\end{equation}%
Eq. (\ref{30}) constitutes the main result of this paper and provides the
explicit dependence on the electric properties of solute and solvent, in
terms of the surface charge density $\sigma $, the permittivity $\varepsilon 
$, the Debye length $\lambda $, and the electrolyte Soret and Seebeck
coefficients $\alpha $ and $\delta \alpha $.

In the case of weak-coupling, $|\hat{\sigma}|\ll 1$, we use $\hat{\zeta}=2%
\hat{\sigma}$, expand (\ref{23}) to quadratic order, $\hat{C}=\hat{\sigma}%
^{2}(1+\alpha +\tau )-6\hat{\sigma}\delta \alpha $, and obtain the transport
coefficient in Debye-H\"{u}ckel approximation, 
\begin{equation}
D_{T}=\frac{\xi e^{2}}{12\eta \varepsilon T}\left( \left( 1+\alpha +\tau
\right) \sigma ^{2}\lambda ^{2}-\delta \alpha \frac{3\sigma \lambda }{\pi
\ell _{B}}\right) .  \label{32}
\end{equation}%
The first term proportional to$\ 1+\alpha +\tau $ agrees with that obtained
previously in \cite{Ras07}. For $\alpha =0=\delta \alpha $ and in the limit $%
\lambda /a\rightarrow 0$, our Eq. (\ref{30}) agrees with the result of \cite%
{Mor99}, and Eq. (\ref{32}) confirms the law $D_{T}\propto \lambda ^{2}$
obtained in Refs. \cite{Ruc81,Pia02,Par04,Fay08}. A linear variation occurs
for small particles \cite{Ras07,Bri03,Fay05,Dho07}, i.e., in the limit
opposite to that treated here. 
\begin{table}[tbp]
\caption{Coupling parameter $\hat{\protect\sigma}=2\protect\pi \protect%
\sigma \protect\lambda \ell _{B}$ calculated from the experimental
parameters of several systems. }%
\begin{tabular}{|l|c|c|c|c|}
\hline
& $\lambda $/nm & $\sigma $/nm$^{-2}$ & $\zeta $/mV & $\hat{\sigma}$ \\ 
\hline
Protein T4L \cite{Put07} &  &  & $26...67$ & $0.5...1.8$ \\ \hline
PS beads \cite{Put07} &  &  & $-85...96$ & $-3$ \\ \hline
SDS micelles \cite{Pia02} & $0.5...2.5$ & $\sim -0.2$ &  & $-0.6...3.1$ \\ 
\hline
Ludox particles \cite{Nin08} & $0.5...8$ & $-0.04$ &  & $-0.1...1.4$ \\ 
\hline
\end{tabular}%
\end{table}

\textit{Discussion.} The main result of the present work concerns the effect
of the thermoelectric field. In the absence of electrolyte Soret and Seebeck
effects ($\alpha =0=\delta \alpha $), the coefficient $D_{T}$ is strictly
positive, i.e., a temperature gradient drives the suspended particles
towards colder regions. An inverse effect ($D_{T}<0$) occurs for a
sufficiently negative Soret coefficient $\alpha $, or if the product $\delta
\alpha \hat{\zeta}$ takes a positive value. In physical terms, $\alpha <0$
means a higher salinity in warmer regions of the solution, whereas $\delta
\alpha $ describes the direction and magnitude of the thermoelectric field
with respect to the thermal gradient.

The numbers of Table I and Ref.\ \cite{Aga89} suggest that protons are the
main source of the thermoelectric effect. The crucial role of the
electrolyte composition is confirmed by the experimental observation that pH
and the presence of protonated buffers significantly influence
thermophoresis \cite{Iac03,Iac06,Put05,Duh06a,Put07,Bra08,Nin08}.\ This is
illustrated in Fig. 1 for NaCl/NaOH solution with different content of the
strong base sodium hydroxide. Sodium chloride has positive Soret and Seebeck
coefficients, thus a slightly negative $D_{T}$ occurs for positively charged
colloids. The large Soret strength of OH$^{-}$ results in $\delta \alpha <0$%
; then the thermoelectric field $\mathbf{E}_{\infty }$ is opposite to the
thermal gradient and drives a negatively charged colloidal particle to
higher $T$ ($D_{T}<0$).

For NaCl solution we compare the Gouy-Chapman or strong-coupling expression (%
\ref{30}) and the Debye-H\"{u}ckel approximation (\ref{32}); according to
the curves in Fig. 1, the latter works well for $|\hat{\sigma}|<\frac{1}{2}$%
, but ceases to be valid at $|\hat{\sigma}|\sim 1$. Most experimental
systems carry rather high charge and surface potential $\zeta =\hat{\zeta}%
k_{B}T/e$, thus requiring a strong-coupling description; the numbers for $%
\hat{\sigma}$ in Table II imply that Debye-H\"{u}ckel approximation fails
for these systems.

Fig. 2 illustrates the effect of the electrolyte composition at low acidity
as a function of the added amount of NaCl or LiCl. The points present
experimental data from Ref. \cite{Put05} for 26-nm polystyrene beads in a
CAPS buffered electrolyte solution. Since the Soret parameters for the
buffer molecules are not known, only Na, Li, Cl, OH are taken into account,
with the values of table 1. At low salinity the thermophoretic mobility is
to a large extent determined by the thermoelectric field of hydroxide ions
and takes a large negative value. Adding salt weakens this effect through
the decreasing relative weight of $\alpha _{\text{OH}}$ in the coefficients $%
\alpha $ and $\delta \alpha $.\ For $n_{S}\gg n_{\text{OH}}$, the pH value
becomes irrelevant for the thermophoretic mobility.\

\begin{figure}
\includegraphics[width=\columnwidth]{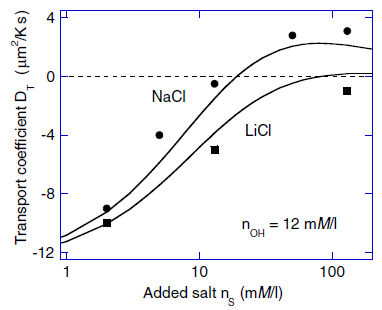}
\caption{Thermophoretic mobility at
large pH.as a function of added salt concentration (NaCl or LiCl). The data
points are taken from Fig. 5a and 5b of Ref. \protect\cite{Put05}; they are
obtained for polystyrene beads of radius $a=13$ nm in a CAPS buffered
electrolyte solution at fixed pH. The curves are calculated from Eq. (%
\protect\ref{30}) with the ionic Soret coefficients of Table 1 for Na or Li,
Cl, and OH, assuming a constant charge density $\protect\sigma =-0.12$ nm$%
^{-2}$. The hydroxide concentration $n_{\text{OH}}=12$ mMol/l corresponds to
pH$=10.3$. }
\end{figure}

We conclude with a discussion of the temperature dependence of $D_{T}$.\ The
Soret coefficients of the alkali chloride serie show a slope $d\alpha
/dT=0.03$ K$^{-1}$ \cite{Cal73,Cal81,Gae82}. Assuming the same law to hold
for $\delta \alpha (T)$ and using the values of Table 1 at 25 $%
{{}^\circ}%
$C, we obtain a good fit for the data of \cite{Iac06} on polystyrene beads
in a 4 mMol/l NaCl solution in the range from 0 to 40 $%
{{}^\circ}%
$C, and in particular, the change of sign of $D_{T}$ at $T=5$ $%
{{}^\circ}%
$C. For comparison, the Debye length $\lambda \sim \sqrt{T\varepsilon }$ and
the permittivity $d\tau /dT<0.01\ $K$^{-1}$ depend weakly on $T$; the
viscosity $d\ln \eta /dT\sim -0.02$ K$^{-1}$ \cite{CRC55} provides an
overall factor to $D_{T}$ but does not affect its sign. The temperature
variation of the ionic Soret coefficients is strongly correlated with the
thermal expansivity $\beta $ of the solvent \cite{Cal73}. In addition to the
electrostatic term, the van der Waals interaction could contribute to $D_{T}$
a term proportional to $\beta $ \cite{Iac06,Bre06}, with a temperature
dependence similar to that of $\alpha $.


\begin{thebibliography}{99}
\bibitem{Duk74} S.S. Dukhin, B.V.\ Derjaguin, in: E. Matijevic (Ed.) \textit{%
Surface and Colloid Science Vol 7}, Wiley New York, (1974)

\bibitem{And89} J.$\ $L.$\ $Anderson, Ann. Rev. Fluid Mech. \textbf{21}, 61
(1989)

\bibitem{Vio00} J.L.\ Viovy, Rev.\ Mod.\ Phys. \textbf{72}, 813 (2000)

\bibitem{Sto04} H.A.\ Stone et al., Ann.\ Rev.\ Fluid Mech. \textbf{36}, 381
(2004)

\bibitem{Wie01} W. K\"{o}hler, S. Wiegand (eds.): \textit{Thermal
Nonequilibrium Phenomena in Fluid Mixtures}, Springer (2001)

\bibitem{Pia02} R.~Piazza, A.~Guarino, Phys. Rev. Lett. \textbf{88}, 208302
(2002)

\bibitem{Dem04} G. Demouchy et al., J. Phys. D: Appl. Phys. \textbf{37},
1417 (2004)

\bibitem{Iac03} S.$\ $Iacopini, R.$\ $Piazza, Europhys. Lett. \textbf{63},
247 (2003)

\bibitem{Put07} S.A. Putnam et al., Langmuir \textbf{23}, 9221 (2007)

\bibitem{Put05} S.A. Putnam, D.G.\ Cahill., Langmuir \textbf{21}, 5317 (2005)

\bibitem{Iac06} S.\ Iacopini et al., EPJ E \textbf{19}, 59 (2006)

\bibitem{Bra08} M.\ Braibanti et al., Phys. Rev. Lett. \textbf{100}, 108303
(2007)

\bibitem{Duh06a} S.\ Duhr, D.\ Braun, PNAS \textbf{103}, 19678 (2006)

\bibitem{Nin08} H. Ning et al., Langmuir \textbf{24}, 2426 (2008)

\bibitem{Gro62} S.R. de Groot, P.\ Mazur, \textit{Non-equlibrium
Thermodynamics}, North Holland Publishing, Amsterdam (1962)

\bibitem{Ruc81} E.$\ $Ruckenstein, J. Colloid Interface Sci. \textbf{83}, 77
(1981)

\bibitem{Par04} A.\ Parola, R.\ Piazza, EPJ E \textbf{15}, 255 (2004)

\bibitem{Fay08} S.\ Fayolle et al., \ Phys. Rev. E \textbf{77} (2008)

\bibitem{Mor99} K.I.\ Morozov, JETP \textbf{88}, 944 (1999)

\bibitem{Wue07} A.$\ $W\"{u}rger, Phys. Rev. Lett. \textbf{98}, 138301 (2007)

\bibitem{Ras07} S.N.\ Rasuli, R. Golestanian, arXiv:0708.0090v1 (2007)

\bibitem{Eas28} E.D.\ Eastman, J.\ Am. Chem. Soc. \textbf{50}, 283 and 292
(1928)

\bibitem{Gut49} G. Guthrie et al., J.\ Chem. Phys.\ \textbf{17}, 310 (1949)

\bibitem{Aga89} J.N. Agar et al., J.\ Phys.\ Chem. \textbf{93}, 2082 (1989)

\bibitem{Cal73} D.R.\ Caldwell, J.\ Phys.\ Chem. \textbf{77}, 2004 (1973)

\bibitem{Cal81} D.R.\ Caldwell, S.A.\ Eide, Deep Sea Res. \textbf{28A}, 1605
(1981)

\bibitem{Gae82} F.S.\ Gaeta et al., J.\ Phys.\ Chem. \textbf{26}, 2967 (1982)

\bibitem{Lan83} L.$\ $D.$\ $Landau, E.$\ $M.$\ $Lifshitz, \textit{%
Electrodynamics of Continuous Media}, Elsevier (1987)

\bibitem{Hie97} P.C. Hiemenz, R.\ Rajagopalan, \textit{Principles of Colloid
and Surface Chemistry}, Dekker (1997)

\bibitem{Bri03} E.\ Bringuier, A.\ Bourdon, PRE \textbf{67}, 011404 (2003)

\bibitem{Fay05} S.\ Fayolle et al., Phys. Rev. Lett.\ \textbf{95}, 208301
(2005)

\bibitem{Dho07} J.K.G.\ Dhont et al., Langmuir \textbf{23}, 1674 (2007)

\bibitem{Bre06} H.\ Brenner, Phys. Rev. E\ \textbf{74}, 036306 (2006)

\bibitem{CRC55} CRC \textit{Handbook of Chemistry and Physics}, 55th
edition, CRC Press (1974)
\end{thebibliography}
\end{document}